\documentclass[final,5p,times,twocolumn,number]{elsarticle}

\usepackage{amssymb}
\usepackage{lipsum}
\usepackage{graphicx}
\usepackage{adjustbox}
\usepackage{amsmath}    
\usepackage[colorlinks=true,linkcolor=green]{hyperref}%
\usepackage{supertabular}
\usepackage{float}
\usepackage[T1]{fontenc}
\usepackage{array} 
\usepackage{enumitem}
\setitemize{noitemsep,topsep=0pt,parsep=0pt,partopsep=0pt}
\usepackage{stfloats} 
\usepackage{float}
\usepackage{xcolor}
\usepackage{booktabs}
\usepackage{siunitx}
\usepackage{etoolbox}
\usepackage{subcaption}
\AtBeginDocument{}

\makeatletter
\def\ps@pprintTitle{%
  \let\@oddhead\@empty
  \let\@evenhead\@empty
  \let\@oddfoot\@empty
  \let\@evenfoot\@oddfoot
}
\makeatother

\newcommand{\degree}{$^o$}
\begin{document}

\begin{frontmatter}

\title{FOSSIL: Thermo-mechanical architecture}

\author[ias]{Valentin Sauvage}
\author[ias]{Clémence de Jabrun}
\author[cea]{Sylvain Martin}
\author[ias]{Bruno Maffei}
\author[ias]{Nabila Aghanim}
\author[ias]{Anaïs Besnard}
\author[ias]{Bruno Borgo}
\author[cea]{Ivan Charles}
\author[ias]{Xavier Coulon}
\author[ias]{Sandrine Couturier}
\author[inaf]{Francesco Cuttaia}
\author[cea]{Jean-Marc Duval}
\author[ias]{Morgane Loquet Le Gall}
\author[ucl]{Giorgio Savini}
\author[inaf]{Luca Terenzi}
\author[ucl]{Berend Winter}

\affiliation[ias]{
    organization={Université Paris Saclay, CNRS, Institut d'Astrophysique Spatiale},
    addressline={Building 121}, 
    city={Orsay},
    postcode={91400}, 
    country={France}}

\affiliation[cea]{
    organization={Université Grenoble Alpes, CNRS, CEA, IRIG, DSBT}, 
    city={Grenoble},
    postcode={38000}, 
    country={France}}

\affiliation[ucl]{
    organization={Astrophysics group, Physics and Astronomy department, UCL}, 
    city={London},
    postcode={GowerStreet, WC1E6BT}, 
    country={United Kingdom}}

\affiliation[inaf]{
INAF - Osservatorio di Astrofisica e Scienza dello Spazio (OAS) di Bologna, via P. Gobetti 93\arraybackslash3, I-40129 Bologna, Italy
}

\begin{abstract}
FOSSIL (FTS fOr CMB Spectral diStortIon expLoration) is a proposed ESA M8 mission tailored to measure the spectral distortions of the Cosmic Microwave Background (CMB) with a sensitivity three orders of magnitude beyond the COBE/FIRAS legacy measurement. Achieving this sensitivity demands an extraordinarily challenging cryogenic architecture: the scientific instrument must be maintained at the lowest achievable temperature (4.5~K), while the detector focal plane assembly operates at 50~mK. This paper presents the thermal architecture of the FOSSIL payload and instrument, from the spacecraft service module at 293~K down to the sub-kelvin detector stage. The thermal design draws on heritage from the Planck and ARIEL missions and relies on a staged passive cooling chain comprising a multi-layer insulation blanket, three V-groove radiators (operating at approximately 130~K, 90~K, and 50~K), and a 25~K actively cooled shield fed by an ESA-provided 4~K cryocooler chain. Sub-kelvin temperatures are achieved via a multi-stage adiabatic demagnetisation refrigerator (ADR) system developed for NewAthena/X-IFU, providing continuous cooling at 1.8~K and 350~mK, and 50~mK with an 80\% duty cycle. The Focal Plane Assembly (FPA), which houses four Kinetic Inductance Detector (KID) arrays at 50~mK, is thermally isolated from the 4.5~K bench via a carbon-fibre reinforced polymer (CFRP) hexapod structure with staged heat interception. We present the steady-state thermal budget across all stages, demonstrating comfortable margins at every temperature level, and discuss key thermal design drivers on the Blackbody Internal Reference (BBIR).
\end{abstract}

\end{frontmatter}


\section{Introduction}

The past three decades of space-based astrophysics have been marked by a steady push toward ever-lower operating temperatures. Missions such as Herschel~\cite{Pilbratt2010}, Planck~\cite{Planck_thermal2011}, SPICA~\cite{Saijo2021_SPICA_CryogenicCooling}, and the forthcoming NewAthena~\cite{Barret2023} and LiteBIRD~\cite{2020JLTP..199..730D} have established that active cryogenic systems combining passive radiative cooling, mechanical cryocoolers, and sub-kelvin refrigerators, are essential to reach the sensitivity levels demanded by modern astrophysical science cases. Each of these missions has contributed to a growing heritage of cryogenic design solutions: V-groove passive radiators demonstrated on \textit{Planck}, multi-stage adiabatic demagnetisation refrigerators (ADRs) developed for NewAthena/X-IFU and LiteBIRD, and dilution refrigerators flown on Planck-HFI. This accumulated heritage now provides a solid technological foundation upon which next-generation missions can build.

CMB science in particular places some of the most stringent cryogenic requirements of any astrophysical discipline. Measuring the faint spectral distortions of the CMB requires not only ultra-sensitive detectors that require to be operated at 50~mK, but also an instrument cold enough (present study shows that 4.5~K is achievable) to suppress its own thermal emission to a level manageable by calibration, and an absolute calibration reference stabilised at temperatures bracketing the CMB monopole at $T_0 \simeq 2.7255$~K~\cite{Fixsen2009}. No existing or previously flown mission has simultaneously addressed all three of these requirements.

The measurement of the absolute frequency spectrum of the Cosmic Microwave Background (CMB) represents one of the most powerful probes of early-Universe physics~\cite{Fixsen1996, Voyage2050SDWP}. Since the landmark COBE/FIRAS measurement~\cite{Fixsen1996, Fixsen2009} and their recent re-analysis~\cite{fabbian2025newconstraintydistortionfiras}, the CMB spectrum has been known to be an exquisitely precise blackbody. However, a broad class of physical processes, from the dissipation of primordial density perturbations and dark matter interactions to the integrated thermal history of large-scale structure formation, inevitably imprint small but characteristic deviations from a pure Planck spectrum, known as spectral distortions~\cite{Sunyaev1970a, Sunyaev1970b, cyr_paper_sd}. These distortions remain undetected, lying three orders of magnitude below the FIRAS sensitivity floor, and constitute a unique guaranteed science target for the next generation of CMB spectrometers.

FOSSIL (FTS fOr CMB Spectral diStortIon expLoration) is a mission concept submitted in response to the ESA M8 call for proposals~\cite{fossil_general_paper}, 
and draws on the design heritage proposals of PIXIE~\cite{pixie2011,Kogut2020PIXIE}, PRISTINE and FOSSIL2022. It aims to map the full sky from 50~GHz to 2~THz using a Martin-Puplett Fourier Transform Spectrometer (FTS) with two inputs and two outputs, providing 130 spectral bins at a resolution of 15~GHz. The primary science goals focus on detecting the $\mu$-type distortion monopole at $\sigma(\mu) \simeq 1.5 \times 10^{-8}$ and the $y$-type distortion monopole at $\sigma(y) \simeq 5 \times 10^{-9}$~\cite{coulon_paper_sd, fossil_general_paper}.

The detector focal plane must be cooled further, to 50~mK, to ensure that the Kinetic Inductance Detectors (KIDs) operate in the photon-noise-limited regime with the required NEP of $\sim 3 \times 10^{-17}$~W~Hz$^{-1/2}$~\cite{Catalano2020,Baselmans2022}. The Blackbody Internal Reference (BBIR), which serves as the absolute reference, must be temperature-controllable within the range $[2.5~\text{K} \rightarrow 2.9~\text{K}]$ to match the CMB spectral emission. A 20~K region, periodically heated, is being studied to simulate the dust contribution at higher frequencies. This stabilisation enables a null operation against the CMB temperature, providing the most precise determination of the latter. Achieving this extraordinary sensitivity imposes extreme requirements on the cryogenic system. The entire instrument (telescopes, FTS optics and scan mechanism~\cite{2022SPIE12190E..2EC}, blackbody reference, and beam-switching mechanisms) must be cooled to 4.5~K to reduce instrumental self-emission, and residuals manageable by calibration.

This paper describes the complete thermal architecture developed for the FOSSIL payload, from the service module interface at ambient temperature down to the sub-kelvin detector stage.


\section{Thermal Requirements and Design Philosophy}
\label{sec:requirements}

\subsection{Science-Driven Temperature Requirements}

The thermal requirements of FOSSIL flow directly from its science objectives. The fundamental measurement principle is a differential comparison between sky emission and an actively cooled blackbody internal reference using a Fourier Transform Spectrometer (FTS). To ensure that instrumental self-emission is at a manageable level, the optimal instrument temperature is as close as possible to the CMB monopole temperature $T_\mathrm{CMB} \simeq 2.7255$~K. A practical upper limit of 4.5~K is adopted for the instrument stage, enabled by a 4~K mechanical cryocooler currently under development at ESA. The feasibility of cooling the instrument closed to the CMB temperature, using the T2 stage of the multi-stage ADR, is currently under investigation (see Section\ref{sec:subK}).

The detector units must operate at 50~mK for two reasons. First, and most fundamentally, the KID materials must operate well below their superconducting critical temperature $T_c$. For the low frequency (LF) band, covering 50--300~GHz, the absorbing resonators require a critical temperature as low as $\sim$0.7~K, achieved through the proximity effect in a tri-layer Al-Ti-Au thin film~\cite{Catalano2015,Catalano2020}. Operating at 50~mK, i.e. at $T \lesssim T_c/10$, ensures a stable superconducting regime and a well-defined kinetic inductance response across the full band. The high frequency (HF) band detectors (300--2000~GHz), based on NbTiN-coupled aluminium distributed absorbers~\cite{Dabironezare2026}, have a higher $T_c$ and could in principle operate at a less demanding temperature; however, housing all four detector units within a single focal plane assembly makes it both practical and thermally consistent to cool the entire FPA to 50~mK. Second, achieving the photon-noise-limited regime requires that a detector noise equivalent power NEP$_\mathrm{det} \lesssim 3 \times 10^{-17}$~W~Hz$^{-1/2}$ in order to be at least one order of magnitude below the mean photon noise NEP$_\mathrm{photon} \sim 4\text{--}5 \times 10^{-16}$~W~Hz$^{-1/2}$ averaged over the bands. State-of-the-art KIDs have demonstrated NEP$_\mathrm{det} \sim 3 \times 10^{-20}$~W~Hz$^{-1/2}$~\cite{Baselmans2022}, comfortably exceeding this requirement.

\subsection{Overall Architecture Concept}

It should be noted that the architecture presented here represents a viable baseline concept at the current level of maturity of the project, rather than a fully consolidated design. The technical solutions will be revisited and optimized during a \textit{Phase~A study}.

\begin{figure*}
    \includegraphics[width=\textwidth]{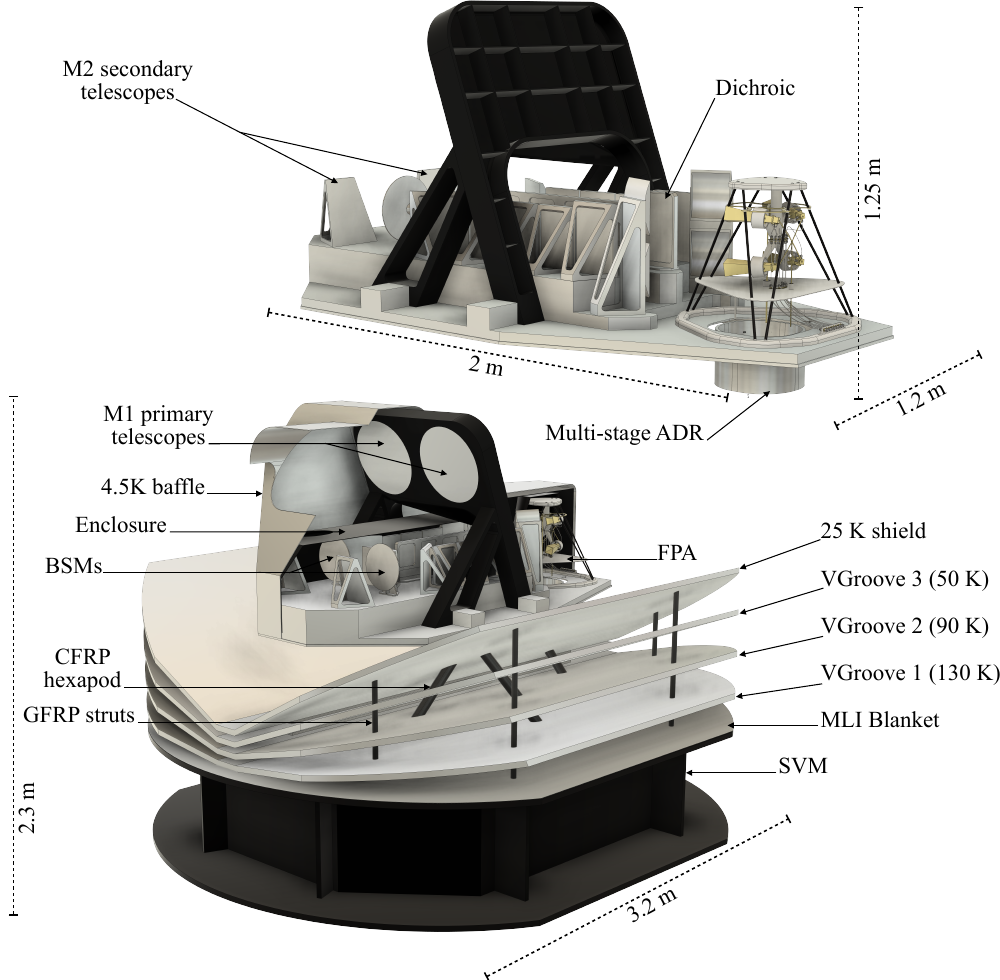}
    
    \caption{\textbf{Bottom}: Overall view of FOSSIL on spacecraft benchmarked and upscaled from ARIEL's. \textbf{Top}: View of FOSSIL's instrument.}
    \label{fig:platform}
\end{figure*}

The thermal architecture of FOSSIL follows a layered approach, with each successive cooling stage progressively reducing the thermal load intercepted by the following stage (Table~\ref{tab:stages}). This staged thermal architecture mirrors the approach successfully employed on ESA's \textit{Planck} mission~\cite{Planck_thermal2011}, with adaptations drawing from the ARIEL satellite platform design~\cite{ARIEL_thermal2022} inspired by the Next Generation Cryogenic IR Telescope (NG-CryoIRTel) study~\cite{NG_CryoIRTel2014}.

\begin{table*}[t]
\centering
\caption{Summary of the FOSSIL thermal components and associated cooling means.}
\label{tab:stages}
\begin{tabular}{p{3cm}p{2.5cm}p{3.5cm}p{6cm}}
\toprule
\textbf{Component} & \textbf{Temperature} & \textbf{Cooling Mean} & \textbf{Primary Function} \\
\midrule
SVM panel       & 293~K            & Passive radiation & Platform, electronics \\
V-Groove 1      & $\sim$130~K       & Passive radiation      & First thermal interception \\
V-Groove 2      & $\sim$90~K        & Passive radiation      & Second thermal interception \\
V-Groove 3      & $\sim$50~K        & Passive radiation      & Third thermal interception \\
25~K shield    & $\sim$25~K        & Active (4~K cooler)   & Fourth thermal interception, LNAs cooling \\
Instrument bench & 4.5~K            & Active (4~K cooler)   & Full instrument \\
BBIR            & 2.5--2.9~K       & ADR T2 stage + heater  & Absolute calibration reference \\
ADR T2 stage    & 1.8~K    & Multi-ADR              & FPA intermediate stage\\
ADR T1 stage    & 350~mK   & Multi-ADR              & FPA intermediate stage \\
FPA / Detectors & 50~mK            & Multi-ADR T0 stage     & KID detector operation \\
\bottomrule
\end{tabular}
\end{table*}

The overall architecture maximises the benefit of the L2 orbit, which provides an exceptionally stable thermal environment, unobstructed by Earth eclipses and with a tightly constrained solar aspect angle (SAA) of approximately $\pm$10\degree. The heritage of \textit{Planck} demonstrates the thermal stability achievable at L2: the HFI bolometer plate was actively stabilised at $\sim$102.8~mK with a double PID regulation system, achieving temperature fluctuations below the required $20~\text{nK}~\text{Hz}^{-1/2}$ in the science frequency band (16~mHz--100~Hz), while the 4~K optical stage was held at 4.81~K with fluctuations meeting the $10~\mu\text{K}~\text{Hz}^{-1/2}$ requirement~\cite{Planck_thermal2011, Pajot2010}. The residual low-frequency fluctuations below 1~mHz were dominated by cosmic ray hits rather than thermal instabilities of the cooling chain, confirming that L2 itself introduces no significant thermal perturbation. This heritage directly informs the FOSSIL thermal architecture, which targets a similar passive stability from the L2 environment while extending active thermal control for the BBIR.

The spacecraft spin axis is aligned quasi-parallel to the Sun-Earth direction, while the instrument line of sight is close to the V-groove vertex direction (see Figure~\ref{fig:platform}), allowing the observations of great circles across the sky while maintaining the V-groove radiators continuously facing deep space. For the purpose of the preliminary thermal model presented in this paper, the following assumptions have been adopted for the SVM thermal design, drawing on solutions demonstrated on \textit{Planck} and ARIEL. The SVM surface facing the Sun is assumed to be covered with solar panel arrays whose cells are selected for a low absorptivity-to-emissivity ratio ($\alpha$/$\epsilon$), with inter-cell gaps treated with optical solar reflectors (OSRs) or second-surface mirrors to maximise specular reflection. Dedicated passive radiators mounted on the lateral SVM faces, coated with OSRs or white paint (e.g.\ AZ-93, $\epsilon > 0.85$, $\alpha < 0.15$), are assumed to supplement the solar panel array in rejecting electronics waste heat toward deep space. Under these assumptions, the SVM panel temperatures are estimated to remain in the range 260--320~K across all operational attitudes within the SAA envelope, ensuring that the interface temperature presented to the MLI blanket at the base of the V-groove stack remains within the design range. These assumptions will be revisited and consolidated during Phase~A in close collaboration with the industrial partners.

\section{From SVM to the instrument: MLI, V-Grooves and 25~K Shield}
\label{sec:passive}

The passive cooling chain of FOSSIL draws directly on the heritage of the \textit{Planck} mission~\cite{Planck_thermal2011}, whose three V-groove radiators and MLI architecture demonstrated the viability of purely passive cooling from the SVM temperature down to $\sim$50~K at L2, with comfortable margins. The FOSSIL design adapts and extends this approach, adding a 25~K actively cooled shield to bridge the gap between the passive chain and the 4.5~K instrument stage. This additional interception stage draws on the design of the SPICA mission~\cite{Saijo2021_SPICA_CryogenicCooling}, which employed a similar actively cooled shield architecture to pre-condition the thermal environment ahead of its sub-kelvin cooling chain. It should be noted that the architecture presented here represents a viable baseline concept at the current level of maturity of the project, rather than a fully consolidated design. Trade-offs on the number of interception stages, their operating temperatures, and the distribution of active versus passive cooling will be revisited and optimised during Phase~A.

\subsection{Multi-Layer Insulation}

The first thermal barrier between the service module (SVM) at $\approx$~293~K and the cryogenic payload is a 20-layer Multi-Layer Insulation (MLI) blanket applied directly to the SVM top surface. With a cylindrical SVM of diameter $D_\mathrm{SVM}$ $\approx$ 3.2~m (calculated from the SAA to ensure a shaded instrument), the MLI blanket covers a top surface area of approximately 8~m$^2$. The effective emissivity of this blanket is $\epsilon_{\mathrm{eff}}$~=~0.008, sufficient to limit the parasitic radiative load transmitted to the first V-groove to a level compatible with the passive cooling budget. This MLI specification is consistent with the \textit{Planck} heritage~\cite{Planck_thermal2011}.

\subsection{V-Groove Passive Radiators}

V-groove passive radiators, arranged in cascade, progressively intercept and re-radiate heat toward deep space. The V-grooves operate as inclined specular reflectors, exploiting their geometry to achieve strong attenuation of the thermal flux between successive stages through multiple reflections toward deep space. The three V-grooves are inclined at angles of 7\degree, 14\degree, and 21\degree with respect to the spacecraft spin axis, operating at equilibrium temperatures of approximately 130 K, 90 K, and 50 K respectively; these values reflect current state-of-the-art performance, and the power budget at each stage indicates that the final temperatures will in fact be lower, to be confirmed through more detailed modelling.

The V-groove panels are modelled as lightweight sandwich structures composed of an aluminium 5056 hexagonal honeycomb core with 0.25~mm thick Al~1085 face sheets, following the ARIEL design heritage~\cite{ARIEL_thermal2022}. The thermal surface treatment strategy, which is critical to the performance of the passive chain, is assumed as follows for the preliminary thermal model:
\begin{itemize}
    \item The \emph{hot face} (toward the SVM) is assumed to be coated with Vapour Deposited Aluminium (VDA), providing high specular reflectivity ($\rho \approx 0.9$) and low IR emissivity, minimising radiative absorption from the warmer stage. These thermo-optical values are consistent with \textit{Planck} and ARIEL heritage~\cite{Planck_thermal2011, ARIEL_thermal2022}.
    \item The \emph{cold face} (facing the colder stage and the instrument) is assumed to exhibit low emittance ($\varepsilon \approx 0.05$), reducing radiative coupling and parasitic heat transfer between successive stages. This value is a modelling assumption within the range achievable with standard space-qualified surface treatments, and will be consolidated against measured data during Phase~A.
    \item Selected surfaces directly exposed to deep space are assumed to be coated with a high emissivity material ($\varepsilon > 0.8$, e.g., MAP PUK or Aeroglaze Z306) to maximize radiative emission toward space.
\end{itemize}
These assumptions are consistent with solutions validated on \textit{Planck}~\cite{Planck_thermal2011} and will be confirmed during Phase~A.

This surface treatment strategy was experimentally validated on the \textit{Planck} mission. The structural support struts connecting the SVM to each V-groove stage are made of GFRP (Glass-Fibre Reinforced Polymer) G-10 material (outer/inner diameter: 26/23~mm), which provides adequate mechanical stiffness with minimised thermal conduction. Heat from these struts is assumed to be fully intercepted at each stage via collars and aluminium braided thermal straps, as a conservative first-order assumption of the preliminary thermal budget.

The small-amplitude Lissajous orbit at L2, with a spin axis quasi-parallel to the Sun-Earth direction ($\pm$ 5\degree) and a nominal Solar Aspect Angle (SAA) of approximately $\pm$10\degree, ensures that the V-groove radiators face deep space continuously. 

\subsection{25~K Actively Cooled Shield}

On top of the three passive V-grooves sits the 25~K shield (inclination angle: 22\degree), actively cooled by the ESA-provided 4~K cooler chain, which delivers 200~mW of cooling power at approximately 20~K. This shield serves two purposes: it provides a further thermal interception stage for parasitic loads (radiations from the 50~K V-groove, conduction from the CFRP hexapod, harness, ...) before they reach the 4.5~K instrument, and it thermally pre-conditions the environment for the Low Noise Amplifiers (LNAs) of the KID readout electronics, which operate between 4~K and 30~K and are mounted on this shield (10~mW heat dissipation each at 20~K).

The structural support between the SVM, the 25~K shield and the 4.5~K instrument bench is provided by a hexapod of six CFRP (Carbon-Fibre Reinforced Polymer) T700 struts (outer/inner diameter: 48/44~mm). This hexapod, sharing the same heritage as the ARIEL payload support structure, provides the dominant conductive heat path due to its larger cross-section compared to the V-groove GFRP struts. Heat intercepted at each stage is handled by collar-and-braid assemblies; a splitting approach with aluminium fittings would ensure total thermal interception without mechanical compromise.


\section{The 4.5~K Instrument Stage}
\label{sec:4K}

\subsection{Instrument Bench and Baffle}

The FOSSIL instrument is mounted on a 4.5~K aluminium bench and surrounded by a 4.5~K light-tight baffle enclosure. The bench houses all optical and quasi-optical elements of the payload: the two off-axis dual-mirror telescopes (primary mirror diameter 420~mm, secondary 200~mm)~\cite{loquetlegall_paper_optics}, the Martin-Puplett FTS with its four wire-grid polarisers and scanning mirror mechanism (FTSM), the blackbody internal reference (BBIR), the two beam-switching mechanisms (BSMs), the dichroic beam splitters, and the focal plane assembly. Cooling the full enclosure to 4.5~K minimises the instrumental self-emission, bringing it close to the CMB monopole temperature and thus reducing the differential signal between the instrument and the sky to a level manageable by the BBIR-based calibration scheme described in Section~\ref{sec:bbir}.

The instrument bench is supported by the hexapod structure from the SVM, as described in Section~\ref{sec:passive}. The cold finger(s) from the 4~K cryocooler provide the thermal interface to the bench, extracting the heat deposited by all internal dissipators (FTS mechanism, BSMs, harnesses) as well as residual parasitic loads from the support structure and baffle.

In the current baseline, the 4.5~K baffle enveloping the instrument is assumed to be made of 2--4~mm thick Al-6061-T6, selected for its high thermal conductivity at cryogenic temperatures and its heritage in space cryogenic instruments. The required thermo-optical properties are: low emissivity ($\varepsilon < 0.1$) on the bottom surface (aluminium-polished) to avoid a thermal radiation coupling with the 25~K shield, high emissivity on the outer surface ($\varepsilon > 0.8$) for self-cooling, and a black-coated inner surface to minimise stray light. The exact thickness and geometry will be consolidated during Phase~A, pending mechanical analysis. The 4.5~K baffle provides an additional safety margin: even for minor pointing excursions, up to the temporary SAA limit of $\pm$ 12\degree (when intermediate V-grooves may receive some transient solar illumination), the 4.5~K baffle itself is not exposed to direct solar radiation. The forbidden region begins at $\pm$27\degree SAA.

\subsection{Heat Sources at 4.5~K}

The dominant heat sources at the 4.5~K stage are:
\begin{itemize}
    \item \textbf{FTS scanning mirror mechanism}: Based on a flexure-mounted reaction-less 4-bar linkage with a custom moving magnet actuator; contributing less than 1.5~mW at 4.5~K during scanning operations (heat dissipation and harness conduction). This low dissipation is achieved through the stiffness compensation design, which reduces the drive current and hence ohmic heating~\cite{Cournoyer2023}.
    \item \textbf{Beam-switching mechanisms (BSMs)}: The transition between the four observatory modes is enabled by two rotation wheels. Each wheel is driven by a cryogenic brushless stepper motor, operated intermittently rather than continuously, with an active power dissipation of approximately 1.5~mW per mechanism. This design builds on heritage from the cryogenic ISOPHOT filter wheels flown on ISO (operated between 290~K and 2.4~K)~\cite{Bollinger1999}, and from wheel mechanisms flown on XMM-Newton and EXOMARS/Pancam (TRL$\,\geq\,$6).
    \item \textbf{ADR heat sink}: During ADR re-magnetisation cycles, the peak parasitic load on the 4~K stage reaches approximately 16~mW (see Section~\ref{sec:subK}).
    \item \textbf{Harnesses and optical fibres}: Heat conducted through harnesses from the SVM to the instrument contribute approximately 3~mW; optical fibres for laser metrology are shielded with ultra-low photon leakage ($<1$~pW at 4~K).
    \item \textbf{Parasitic loads}: A radiative load of 9~mW and a conductive load of 8~mW, originating from the 25~K actively cooled shield via radiative coupling and conduction through the CFRP support structure respectively, are intercepted at the 4.5~K instrument baffle. 
\end{itemize}

The total maximum load at the 4.5~K stage is approximately 39~mW, against a 4~K cooler capacity of 50~mW at 4.5~K~\cite{ESA_M8F3_Call2025}, providing a 28\% margin. This figure conservatively includes the 16~mW ADR heat-sink load, which is a peak value reached only during the T0 stage re-magnetisation, as well as the BSM contribution, which occurs only when the mechanisms are operated. Excluding the parasitic loads intercepted at the baffle (17~mW), the internal dissipators alone amount to 22~mW.

\subsection{Telescope and Optical Elements}

Aluminium is assumed as the baseline material for all mirrors, providing excellent thermal uniformity at cryogenic temperatures and a uniform coefficient of thermal expansion (CTE) across the full optical bench, avoiding the risk of thermo-mechanical misalignment during cool-down.

Every optical surface adds an emission term $\epsilon(\nu, T)·B_ν(T)$ to the measured signal, so that each mirror contributes its own spectral signature to the final spectrum. Since the emissivity depends on both temperature and frequency, no single value can describe this contribution: a dedicated characterisation of each optical elements, over the temperature and frequency ranges relevant to FOSSIL, will be required to subtract these instrumental terms from the final spectra.


\section{Sub-Kelvin Cooling Chain: Multi-Stage ADR}
\label{sec:subK}

\subsection{ADR System Overview}

Cooling below 4.5~K is provided by a multi-stage Adiabatic Demagnetisation Refrigerator (ADR) system derived from the design being developed for the NewAthena/X-IFU instrument~\cite{Prouve2020,ADR_SPIE_2026}. This cooler also benefits from developments for SPICA/SAFARI~\cite{DUBAND2014213} and LiteBIRD~\cite{2020JLTP..199..730D} studies.

The ADR system provides three successive cooling stages:
\begin{itemize}
    \item \textbf{T2 continuous stage (4.5~K $\rightarrow$ 1.8~K)}: Provides continuous cooling power of \textbf{1~mW at 1.8~K}. This stage intercepts the dominant parasitic loads from the FPA support structure and FPA shield, and provides the cooling power to maintain the BBIR within the 2.5~--~2.9~K range.
    \item \textbf{T1 continuous stage (1.8~K $\rightarrow$ 0.35~K)}: Provides continuous cooling power of \textbf{10~$\mu$W at 350~mK}, dedicated exclusively to intermediate thermal interception within the FPA hexapod.
    \item \textbf{T0 single-shot stage ($\rightarrow$ 50~mK)}: Provides cooling power of \textbf{0.86~$\mu$W at 50~mK}, with a hold time exceeding 80\% of a 35-hour cycle, i.e. more than 28~hours of continuous science operations. This stage directly cools the detector units and feedhorns via a copper thermal link. Based on the temperature sensitivity of the aluminium KIDs, the maximum allowable temperature fluctuation at the 50~mK stage is $\delta T \lesssim 50$~$\mu$K~Hz$^{-1/2}$. This is comfortably compatible with the demonstrated ADR thermal stability of $<0.4$~$\mu$K~Hz$^{-1/2}$ achieved on the NewAthena/X-IFU demonstrator~\cite{Maisonnave2022_ADR}, confirming that the FOSSIL ADR heritage provides ample thermal stability margin for KID operation.
\end{itemize}

\subsection{ADR Operational Cycle}

ESA margin requirements impose a 100\% margin at the 50~mK stage, meaning that the ADR T0 stage is sized to deliver twice the expected load, leaving 50\% of its cooling capacity in reserve throughout the science operations. This reserve provides operational flexibility: the in-flight scanning strategy and ADR cycling will be tailored to maximise scientific return within the available thermal budget.

During the re-magnetisation mode, which lasts approximately 7 hours, the ADR power consumption rises from 30~W to 69~W (peak). The heat deposited at the 4~K heat sink during this peak demand drives the 4~K cooler to its maximum capacity. The T2 and T1 stages continue to operate continuously throughout the mission duration.


\section{Focal Plane Assembly Thermal Design}
\label{sec:fpa}

\subsection{FPA Architecture}

The Focal Plane Assembly (FPA), whose multi-stage thermal isolation draws on the philosophy demonstrated on \textit{Planck} HFI~\cite{Planck_thermal2011} and more recent structural and thermal analyses~\cite{Sauvage2025_StructuralThermalCCDR}, houses four detector units (two Low-Frequency Detector Units, LFDUs, covering 50--300~GHz, and two High-Frequency Detector Units, HFDUs, covering 300--2000~GHz). Each detector unit consists of a smoothwall multimoded feedhorn coupled to a segmented KID array (baseline: $4 \times 4$ pixels, to be confirmed during Phase~A). The FPA is built around an isostatic hexapod structure designed to support approximately 5--7~kg, surrounded by a 1.8~K radiation shield (described in Section~\ref{sec:subK}). The total height of the FPA is approximately 500~mm.

In the current baseline, the FPA shield is assumed to be made of 2~mm Al-6061-T6 with an electrodeposition of Niobium on the inside for electromagnetic shielding, with openings for beam illumination of the feedhorns. The whole shield is blackened to suppress stray light. The detector units are assumed to be supported by an Al-6061-T6 interface at a mean temperature of $\sim$150~mK, pending mechanical consolidation during Phase~A.

\subsection{Multi-Stage Thermal Isolation}

\begin{figure}[htbp]
    \centering

    \begin{subfigure}{0.5\textwidth}
        \centering
        \includegraphics[width=\textwidth, keepaspectratio]{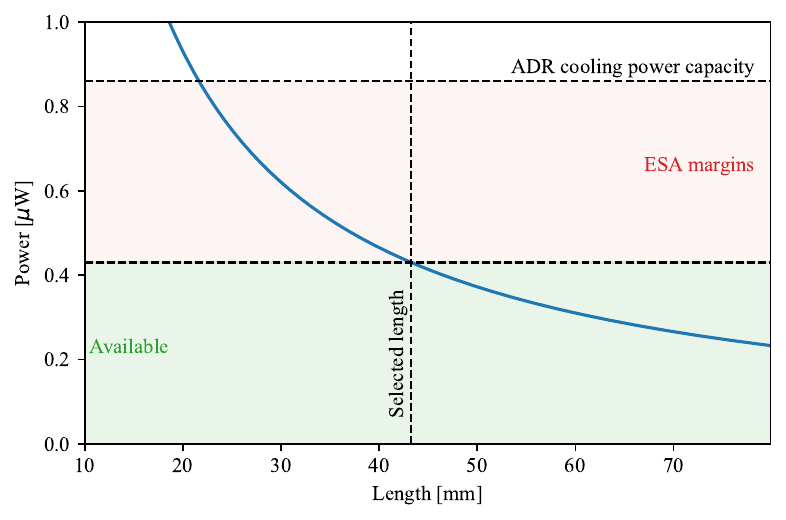}
        \caption{}
        \label{fig:tradeoff_length_350_50_FPA}
    \end{subfigure}

    \begin{subfigure}{0.5\textwidth}
        \centering
        \includegraphics[width=\textwidth, keepaspectratio]{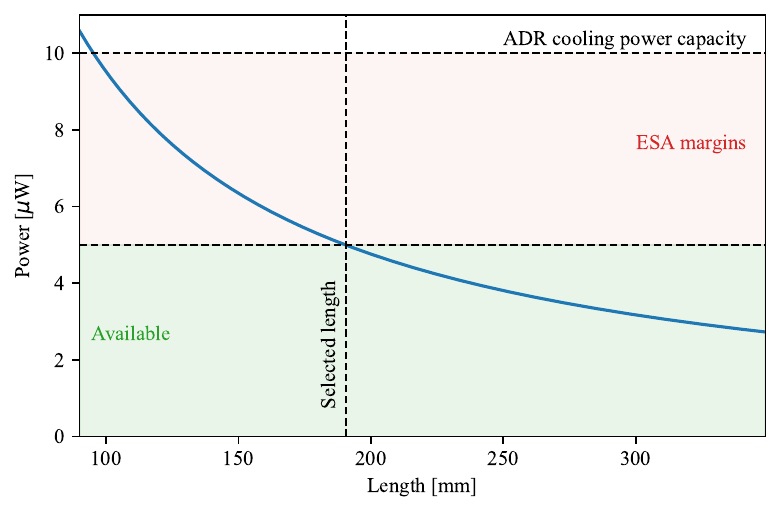}
        \caption{}
        \label{fig:tradeoff_length_cfrp_1800_350_FPA}
    \end{subfigure}

    \begin{subfigure}{0.5\textwidth}
        \centering
        \includegraphics[width=\textwidth, keepaspectratio]{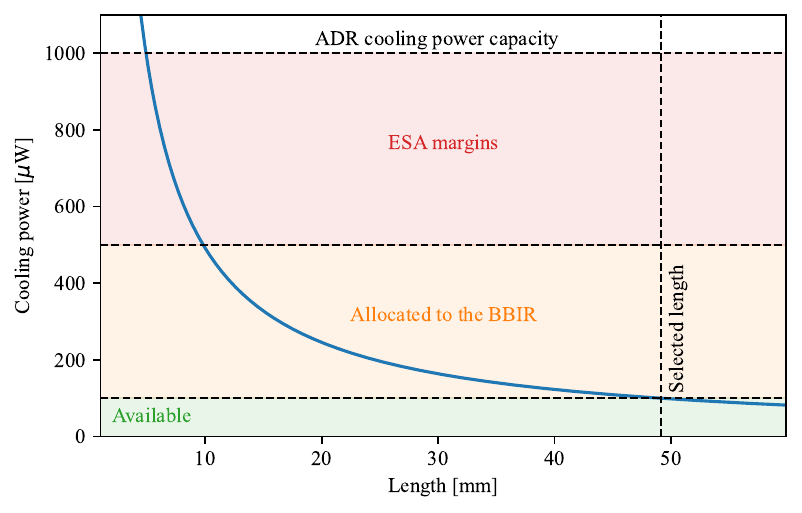}
        \caption{}
        \label{fig:tradeoff_length_cfrp_4500_1800_FPA}
    \end{subfigure}

    \caption{Trade-off between the parasitic heat load from the CFRP struts length and the available cooling power (with ESA margins) for the three temperature stages: (a) 350--50 mK, (b) 1.8--350 mK, and (c) 4.5--1.8 K.}
    \label{fig:tradeoff_length_FPA}
\end{figure}

Maintaining all four detector units at 50~mK while the bench is at 4.5~K, with a maximum conductive load budget of 440~nW at this stage, requires very effective thermal isolation. A multi-stage heat interception scheme is employed:

\begin{enumerate}
    \item \textbf{CFRP struts (T700 grade)}: The primary structural supports between the 4.5~K bench and the FPA 1.8~K interface rely on CFRP struts (outer/inner diameter: 7.8/6.8~mm). These dimensions result from a compromise between the mechanical rigidity required to support the FPA mass and the maximum admissible conductive heat load at the 1.8~K stage (see Table~\ref{tab:fpa_budget}), taking advantage of CFRP's low thermal conductivity and excellent strength-to-weight ratio. The conductive heat load through a single CFRP strut between two temperature stages $T_\mathrm{hot}$ and $T_\mathrm{cold}$ is estimated as:
\begin{equation}
\dot{Q}_\mathrm{cond} = \frac{S}{L} \int_{T_\mathrm{cold}}^{T_\mathrm{hot}} \kappa(T)\, \mathrm{d}T
\label{eq:conduction}
\end{equation}
where $S = \frac{\pi}{4} (D_\mathrm{out}^2 - D_\mathrm{in}^2)$ is the cross-sectional area of the hollow tube ($D_\mathrm{out} = 7.8$~mm, $D_\mathrm{in} = 6.8$~mm), $L = 33$~mm is the strut length, and $\kappa(T)$ is the temperature-dependent thermal conductivity of CFRP T700, integrated between the two intercept temperatures. For CFRP T700, the integrated conductivity is derived from dedicated cryogenic measurements~\cite{sauvage:2026}, yielding a conductive load of approximately 72~nW per strut for the 350~mK--50~mK section (i.e. $\sim$430~nW for the six struts), consistent with the 50~mK budget of Table~\ref{tab:fpa_budget}.

Each strut must simultaneously satisfy the strength and stability criteria under the launch load. The design load per strut is $F_\mathrm{design} = 1812$~N, derived from the FPA mass of 7~kg combined with the launcher quasi-static load requirement (100~g acceleration). The compressive stress in the strut wall, $\sigma = F_\mathrm{design}/S = 158$~MPa, compared to the T700 composite strength of $\sigma_\mathrm{ult} \simeq 2550$~MPa~\cite{Toray_T700S}, yields a material (crushing) safety factor of 16. In addition, for a hollow cylindrical tube with fixed-fixed boundary conditions ($K = 0.5$, taken as $K = 1$ to remain conservative), the Euler critical buckling load is:
\begin{equation}
F_\mathrm{cr} = \frac{\pi^2 E I}{L^2} \geq F_\mathrm{design}
\label{eq:buckling}
\end{equation}
where $E = 70$~GPa is the Young's modulus of the CFRP T700 laminate and $I = \frac{\pi}{4}(r_\mathrm{out}^4 - r_\mathrm{in}^4) = 76.7$~mm$^4$ is the second moment of area of the tube cross-section. With the baseline dimensions (outer/inner diameter: 7.8/6.8~mm) and a strut length of 33~mm, the critical buckling load is $F_\mathrm{cr} \simeq 48.7$~kN, corresponding to a buckling safety factor of $\sim$27: the strut sizing is therefore driven by material strength rather than by buckling.

    \item \textbf{Staged interception at 1.8~K and 350~mK}: The CFRP struts are split at the 1.8~K and 350~mK stages, with thermal collars and aluminium or copper braided straps at each intercept point. This splitting approach prevents heat from propagating continuously from 4.5~K to 50~mK through the structural path. The trade-off length VS cooling power, taking ESA margins into consideration, is represented in Figure~\ref{fig:tradeoff_length_FPA}.

    \item \textbf{Passive isolation below 350~mK}: Two complementary phenomena provide additional passive isolation at the coldest stages. First, all structural STM interface at focal plane level use Al-6061-T6, which becomes superconducting below 1.1~K; below this transition, its thermal conductivity drops dramatically, providing passive isolation at both the 350~mK and 50~mK stages. Second, at the 350~mK--50~mK boundary, the Kapitza boundary resistance provides further passive thermal isolation, reducing parasitic conductive loads on the coldest ADR stage.
\end{enumerate}

\begin{figure*}[h]
\centering
\includegraphics{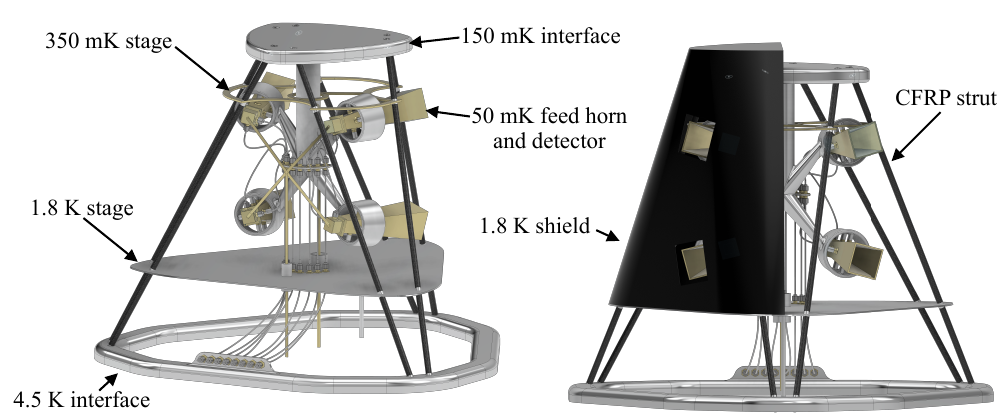}
\caption{CAD model of the FOSSIL Focal Plane Assembly (FPA). \textit{Left}: Internal view showing the multi-stage thermal architecture, with the CFRP hexapod struts connecting the 4.5~K interface (bottom crown) to the 1.8~K intermediate platform and the 150~mK detector interface (top plate). The 50~mK feed horns and detector units are visible at the center, surrounded by the 350~mK interception stage. \textit{Right}: External view with the 1.8~K radiation shield in place, enclosing the cold stages and limiting the radiative load from the 4.5~K environment on the sub-kelvin components. The CFRP T700 struts (outer/inner diameter: 7.8/6.8~mm) are split at the 1.8~K and 350~mK stages to prevent continuous conductive heat propagation from the 4.5~K bench to the 50~mK detector stage.}
\label{fig:stm_fpa}
\end{figure*}

\subsection{Thermal Straps}

Copper thermal straps (Residual Resistivity Ratio (RRR) = 50) are implemented at the 50~mK, 350~mK, and 1.8~K stages to re-direct heat flows from structural supports to the ADR cold stages. At the 50~mK stage, the copper thermal strap is sized to keep the temperature gradient along its length below 5~mK, corresponding to a computed cross-section-to-length (S/L) ratio of 0.183~mm. At the 1.8~K stage, the strap also intercepts approximately 1~$\mu$W of radiative flux from the 4.5~K environment through the 1.8~K aluminium shield.

\subsection{Thermal Power Balance at Sub-Kelvin Stages}

Table~\ref{tab:fpa_budget} summarises the power balance at each sub-kelvin stage, confirming that all stages operate with positive margins.

\begin{table}[]
\centering
\caption{Sub-kelvin thermal power budget for FOSSIL.}
\label{tab:fpa_budget}
\begin{tabular}{llll}
\toprule
\textbf{Stage} & \textbf{Load} & \textbf{Cooling power} & \textbf{Margin} \\
\midrule
50~mK   & 430~nW  & 860~nW  & 100\% \\
350~mK  & 5~$\mu$W & 10~$\mu$W & 100\% \\
1.8~K   & 500~$\mu$W & 1~mW    & 100\% \\
\bottomrule
\end{tabular}
\end{table}

The 50~mK budget includes contributions from conduction through CFRP struts, residual radiation through the 1.8~K shield, and conduction through the superconducting coaxial readout cables. No electronics are located at the 50~mK stage; heat dissipation from active components is zero at this temperature level.

\section{Thermal Constraints of the Blackbody Internal Reference}
\label{sec:bbir}

\subsection{Role and Thermal Requirements}

In the science observation modes, one FTS input arm is directed at the sky through a telescope, while the other is directed at the BBIR. The BBIR must therefore represent a known, stable, nearly perfect blackbody emission at temperatures bracketing the CMB monopole temperature $T_0 \approx 2.7255$~K, with setpoints spanning $[2.5~\text{K} \rightarrow 2.9~\text{K}]$.

The BBIR assembly comprises two thermally independent regions: a primary absorbing region actively controlled over 2.5--2.9~K, and a secondary high-frequency calibration cavity that can be independently heated to $\sim$20~K during dedicated calibration sequences. 

\subsection{Thermo-mechanical Interface}

The BBIR is thermally isolated from the 4.5~K bench by insulating struts, whose material and cross-section are tuned to keep the conductive heat load below 120~$\mu$W. Combined with the 80~$\mu$W allocated to the 20~K BBIR region, this brings the total BBIR load on the ADR T2 stage below 200~$\mu$W. The structural substrate is Al-6061-T6 to minimise thermal gradients across the absorbing surface. The BBIR aperture is approximately 200~mm, with an estimated mass in the range 5--7.5~kg. The absorbing surface design draws on heritage from the ARCADE experiment~\cite{Fixsen2011} and the MetOp-SG MWI calibration target~\cite{Simonetto2021}.

Cooling of the absorbing region to its 2.5--2.9~K setpoints is provided by the ADR T2 stage (Section~\ref{sec:subK}), via a thermal link sized to allow active temperature control. The feasibility of a second region, periodically heated to 20~K, is envisioned to simulate the high frequency dust contribution. 

\subsection{Temperature Monitoring and Stability}

The BBIR temperature is monitored by thick-film ruthenium oxide (RuO$_2$) thermometers, readable to 0.1~mK precision in 1~s. Uncertainty in the absolute thermometry scale from ground calibration yields $\Delta T_0 \approx 100~\mu$K, sufficient for FOSSIL's spectral distortion measurement goals. The BBIR temperature control system is part of the Temperature Monitoring and Control System (TMCS), developed at IAC/IFCA drawing on LiteBIRD heritage~\cite{2020JLTP..199..730D}.


\section{Full Steady-State Thermal Budget}
\label{sec:budget}

\subsection{Heat Flow Chain}

\begin{figure*}[h]
\centering
\includegraphics{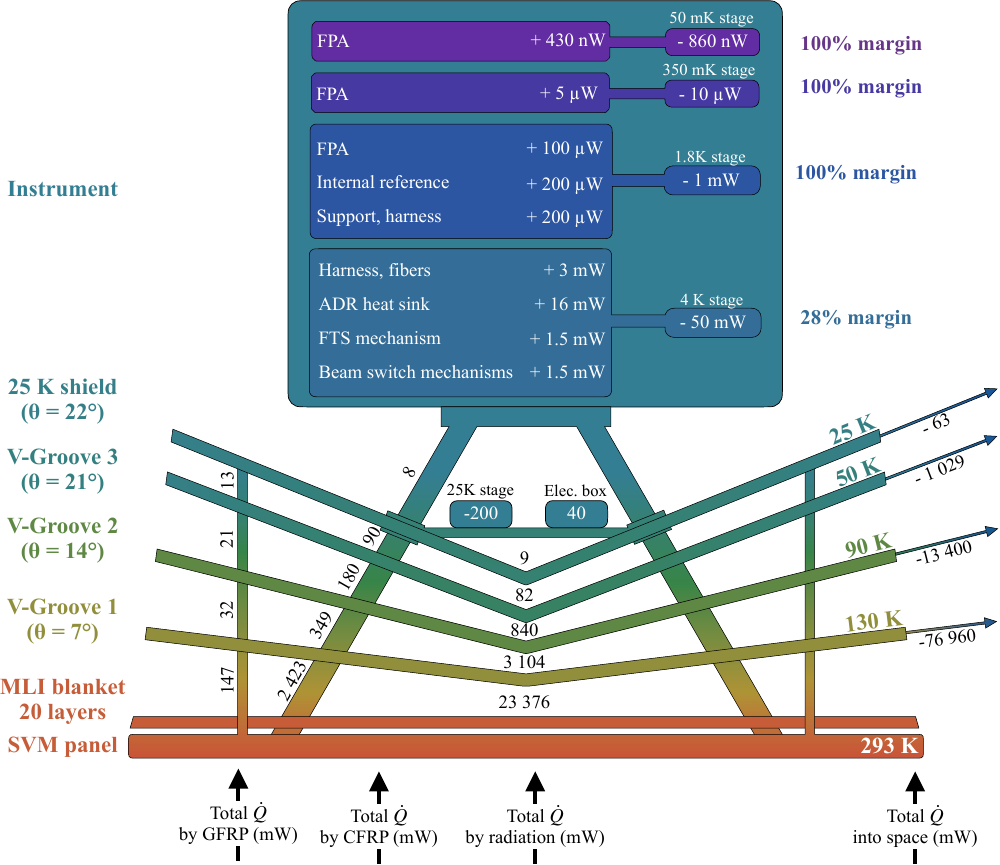}
\caption{Steady-state heat-flow budget from the SVM (293~K) down to the 50~mK stage. Radiative and conductive loads across the 130~K, 90~K, 50~K, and 25~K V-grooves and shield are shown with corresponding surface properties and angles. Structural (CFRP/GFRP), MLI, and radiative contributions are quantified.}
\label{fig:thermal_budget}
\end{figure*}

Figure~\ref{fig:thermal_budget} summarizes the steady-state heat-flow budget from the SVM panel at 293~K to the detector stage, showing the dominant contributions at each level. The figures account for radiative and conductive loads across the 130~K, 90~K, 50~K, and 25~K stages.

\subsection{Passive Cooling Margins}

The preliminary thermal budget, retaining only dominant contributions, provides passive cooling margins of approximately:
\begin{itemize}
    \item $\sim$55~W at 130~K (V-Groove 1)
    \item $\sim$11~W at 90~K (V-Groove 2)
    \item $\sim$100~mW at 50~K (V-Groove 3)
\end{itemize}

These margins can be tuned through geometric adjustments (V-groove surface area, inclination angles) or by adding dedicated radiator panels at the V-groove edges, providing flexibility to meet the 4~K cooler pre-cooling requirements. The thermal contribution of harnesses and optical fibres routed from the SVM to the instrument is negligible relative to the power levels involved at the V-groove stages.

\subsection{Active Cooling Margins}

At the 4.5~K stage, the total instrument load is approximately 39~mW against a 4~K cooler capacity of 50~mW, providing a 28\% margin (Section~\ref{sec:4K}). At the 25~K stage, the dominant incident loads are the radiative (82~mW) and conductive (90~mW through the CFRP hexapod, 13~mW through the GFRP struts) heat fluxes from the 50~K (third) V-groove, complemented by the dissipation of the four LNAs (up to 10~mW each, i.e. 40~mW as a conservative worst case). Of this 225~mW gross incident load, 8~mW are conducted onward to the 4.5~K bench through the CFRP hexapod, 9~mW are radiated to the 4.5~K baffle, and 63~mW are passively re-radiated toward deep space by the space-facing surfaces of the shield, leaving a net load of 145~mW on the 20~K pre-cooler stage against its 200~mW capacity, providing a 38\% margin.

Table~\ref{tab:margins} summarises the margins at all active cooling stages.

\begin{table}
\centering
\caption{Active cooling power budget and margins at each FOSSIL cryogenic stage.}
\label{tab:margins}
\begin{tabular}{lllll}
\toprule
\textbf{Stage} & \textbf{Load} & \textbf{Capacity} & \textbf{Margin} \\
\midrule
25~K     &   145~mW   &  200~mW    & 38~\% \\
4.5~K     &  39~mW   & 50~mW     & 28\%\\
1.8~K   & 500~$\mu$W & 1~mW    & 100\% \\
350~mK  & 7.5~$\mu$W & 10~$\mu$W & 100\% \\
50~mK   & 440~nW  & 860~nW  & 100\% \\
\bottomrule
\end{tabular}
\end{table}


\section{Conclusion}
\label{sec:conclusion}

The FOSSIL thermal architecture builds on a strong heritage of cryogenic space missions while extending established approaches to meet the more demanding requirements of CMB spectroscopy. Planck-HFI represented the previous state-of-the-art space cryogenic chain, achieving 0.1~K with a $^3$He/$^4$He dilution refrigerator and demonstrating the viability of the V-groove passive cooling approach at L2\cite{Planck_thermal2011}. FOSSIL adopts the same passive cooling philosophy together with a 4.5~K pre-cooler stage deriving from Planck, but pushes it further with a detector stage at 50~mK rather than 100~mK. The use of a multi-stage ADR instead of a dilution refrigerator provides a more autonomous and robust path to 50~mK, with a clear technology maturation route through NewAthena/X-IFU. The FOSSIL V-groove and 25~K shield also benefit directly from ARIEL structural heritage~\cite{ARIEL_thermal2022, ariel2018}, including honeycomb sandwich construction and CFRP strut architectures, while spacecraft benchmarking has shown that either an upscaled ARIEL platform or a Planck-heritage SVM could accommodate the payload requirements. The multi-stage ADR further builds on SPICA/SAFARI studies~\cite{Saijo2021_SPICA_CryogenicCooling} and LiteBIRD developments~\cite{2020JLTP..199..730D}; by mid-2026, a demonstration model of the X-IFU multi-ADR system will be delivered to the NewAthena consortium for coupling with the X-IFU FPA, and by the end of FOSSIL's Phase A a TRL-6 engineering model is expected to be available.

The resulting architecture addresses the central thermal challenge of maintaining a complex optical instrument at 4.5~K and its detectors at 50~mK, while preserving a blackbody internal reference near T$_{CMB}$ within the resource constraints of an ESA medium-class mission. This is achieved through a staged passive cooling chain comprising MLI, three V-grooves, and a 25~K active shield, which reduces the thermal environment from 293~K to 25~K with comfortable margins; an ESA-provided 4~K cryocooler delivering 50~mW at 4.5~K and 200~mW at 20~K; and a multi-stage ADR providing continuous cooling at 1.8~K and 350~mK and 50~mK operation with $>$80\% duty cycle, with 100\% margin at 50~mK, 100\% at 350~mK and 100\% at 1.8~K. A CFRP hexapod FPA support with staged heat interception, exploiting the superconducting transition of Al-6061-T6 below 1.1~K and the low thermal conductivity of CFRP, further limits conductive heat loads at the coldest stages. The BBIR is supported by ARCADE and MetOp-SG heritage, while low-dissipation beam-switching mechanisms and the FTS motor reduce the demands placed on the cooling chain.

At every temperature stage, the preliminary thermal budget demonstrates positive margins, confirming the compatibility of the architecture with the mission requirements. The overall design is supported by a coherent technology-development pathway: the multi-stage ADR will reach TRL-6 through NewAthena/X-IFU; KID detectors are being advanced through dedicated development programmes, including CNES-funded CMB spectral-distortion KIDs R\&D, and developments at SRON and Institut Néel; and pathfinder instruments such as BISOU~\cite{bisou2024}, TMS~\cite{TMS2020}, and COSMO~\cite{cosmo2024} will provide additional validation of key subsystems in representative environments. The FOSSIL thermal architecture therefore combines demonstrated cryogenic heritage with targeted technological advances, providing a technically credible and well-grounded basis for the next major advance in CMB spectroscopy and supporting the feasibility of a mission targeted for launch in the early 2040s.


\section*{Acknowledgements}

The authors thank the FOSSIL consortium and the CNES for their efforts and contributions to the mission concept. 


\bibliographystyle{elsarticle-num}
\bibliography{bibliography/literature}

@article{Fixsen1996,
  author  = {Fixsen, D.J. and others},
  title   = {The Cosmic Microwave Background spectrum from the full {COBE} {FIRAS} data set},
  journal = {Astrophys. J.},
  year    = {1996},
  volume  = {473},
  pages   = {576}
}

@article{Fixsen2009,
  author  = {Fixsen, D.J.},
  title   = {The Temperature of the Cosmic Microwave Background},
  journal = {Astrophys. J.},
  year    = {2009},
  volume  = {707},
  pages   = {916--920}
}

@article{Voyage2050SDWP,
  author  = {Chluba, J. and others},
  title   = {New horizons in cosmology with spectral distortions of the cosmic microwave background},
  journal = {Exp. Astron.},
  year    = {2021},
  volume  = {51},
  number  = {3},
  pages   = {1515--1554},
  doi     = {10.1007/s10686-021-09729-5}
}

@article{Sunyaev1970a,
  author  = {Sunyaev, R.A. and Zeldovich, Ya.B.},
  title   = {Small scale entropy and adiabatic density perturbations: Antimatter in the Universe},
  journal = {Astrophys. Space Sci.},
  year    = {1970},
  volume  = {9},
  pages   = {368--382}
}

@article{Sunyaev1970b,
  author  = {Zeldovich, Ya.B. and Sunyaev, R.A.},
  title   = {The Interaction of Matter and Radiation in the Hot Model of the Universe, {II}},
  journal = {Astrophys. Space Sci.},
  year    = {1970},
  volume  = {7},
  pages   = {20--30}
}

@misc{fabbian2025newconstraintydistortionfiras,
      title={A new constraint on the $y$-distortion with FIRAS: implications for feedback models in galaxy formation and cosmic shear measurements}, 
      author={Giulio Fabbian and Federico Bianchini and Alina Sabyr and J. Colin Hill and Christopher C. Lovell and Leander Thiele and David N. Spergel},
      year={2025},
      eprint={2512.03038},
      archivePrefix={arXiv},
      primaryClass={astro-ph.CO},
      url={https://arxiv.org/abs/2512.03038}, 
}

@article{Planck_thermal2011,
  author  = {{Planck Collaboration}},
  title   = {Planck early results. {II}. The thermal performance of {Planck}},
  journal = {Astron. Astrophys.},
  year    = {2011},
  volume  = {536},
  pages   = {A2},
  doi     = {10.1051/0004-6361/201116486}
}

@article{Pajot2010,
  author  = {Pajot, F. and Ade, P.~A.~R. and Beney, J.-L. and others},
  title   = {{Planck pre-launch status: HFI ground calibration}},
  journal = {A\&A},
  year    = {2010},
  volume  = {520},
  pages   = {A10},
  doi     = {10.1051/0004-6361/200913203}
}

@article{Pilbratt2010,
  author  = {Pilbratt, G.~L. and Riedinger, J.~R. and Passvogel, T. and Crone, G. and Doyle, D. and Gageur, U. and Heras, A.~M. and Jewell, C. and Metcalfe, L. and Ott, S. and Schmidt, M.},
  title   = {{Herschel Space Observatory. An ESA facility for far-infrared and submillimetre astronomy}},
  journal = {A\&A},
  year    = {2010},
  volume  = {518},
  pages   = {L1},
  doi     = {10.1051/0004-6361/201014759}
}

@article{ARIEL_thermal2022,
  author  = {Morgante, G. and others},
  title   = {The thermal architecture of the {ESA} {ARIEL} payload at the end of phase {B1}},
  journal = {Exp. Astron.},
  year    = {2022},
  volume  = {53},
  pages   = {905--944},
  doi     = {10.1007/s10686-022-09851-y}
}

@article{ariel2018,
  author  = {Puig, L. and others},
  title   = {The Phase {A} study of the {ESA} {M4} mission candidate {ARIEL}},
  journal = {Exp. Astron.},
  year    = {2018},
  volume  = {46},
  number  = {1},
  pages   = {211--239},
  doi     = {10.1007/s10686-018-9604-3}
}

@inproceedings{Saijo2021_SPICA_CryogenicCooling,
  author    = {Saijo, M. and others},
  title     = {Thermal Design of {SPICA} Cryogenic Cooling System},
  booktitle = {Proc. 50th International Conference on Environmental Systems (ICES)},
  year      = {2021},
  pages     = {108}
}

@misc{ESA_M8F3_Call2025,
  author       = {{ESA Director of Science}},
  title        = {{Call for a Medium-size and a Fast mission opportunity in ESA's Science Programme}},
  year         = {2025},
  howpublished = {ESA document ESA-SCI-DIR-AO-013, \url{https://www.cosmos.esa.int/web/call-for-missions-2025}},
  note         = {Released 19 March 2025; Step-2 M-class deadline 19 March 2026}
}

@article{Barret2023,
  author  = {Barret, D. and Albouys, V. and den Herder, J.-W. and Piro, L. and Cappi, M. and Huovelin, J. and Kelley, R. and Mas-Hesse, J.~M. and Paltani, S. and Rauw, G. and Rozanska, A. and Svoboda, J. and Wilms, J. and Yamasaki, N. and others},
  title   = {{The Athena X-ray Integral Field Unit: a consolidated design for the system requirement review of the preliminary definition phase}},
  journal = {Experimental Astronomy},
  year    = {2023},
  volume  = {55},
  number  = {2},
  pages   = {373--426},
  doi     = {10.1007/s10686-022-09880-7}
}

@techreport{NG_CryoIRTel2014,
  author      = {{European Space Agency}},
  title       = {{NG-CryoIRTel}: {CDF} Study Report ({CDF-152(A)})},
  institution = {ESA},
  year        = {2014},
  type        = {ESA Technical Report}
}

@article{2020JLTP..199..730D,
  author  = {Duval, J.-M. and others},
  title   = {{LiteBIRD} Cryogenic Chain: 100~{mK} Cooling with 
             Mechanical Coolers and {ADRs}},
  journal = {J. Low Temp. Phys.},
  year    = {2020},
  volume  = {199},
  number  = {3--4},
  pages   = {730--736},
  doi     = {10.1007/s10909-020-02371-z}
}

@article{Prouve2020,
  author  = {Prouv{\'e}, T. and Duval, J.-M. and Charles, I. and Yamasaki, N.Y. and Mitsuda, K. and Nakagawa, T. and Shinozaki, K. and Tokoku, C. and Yamamoto, R. and Minami, Y. and Le~Du, M. and Andre, J. and Daniel, C. and Linder, M.},
  title   = {{ATHENA X-IFU 300~K--50~mK cryochain test results}},
  journal = {Cryogenics},
  year    = {2020},
  volume  = {112},
  pages   = {103144},
  doi     = {10.1016/j.cryogenics.2020.103144}
}

@article{Catalano2020,
  author  = {Catalano, A. and others},
  title   = {Sensitivity of {LEKID} for space applications between 80~{GHz} and 600~{GHz}},
  journal = {Astron. Astrophys.},
  year    = {2020},
  volume  = {641},
  pages   = {A179},
  doi     = {10.1051/0004-6361/202038199}
}

@article{Catalano2015,
  author        = {Catalano, A. and Goupy, J. and le~Sueur, H. and others},
  title         = {Bi-layer kinetic inductance detectors for space observations between 80--120~{GHz}},
  journal       = {Astron. Astrophys.},
  year          = {2015},
  volume        = {580},
  pages         = {A15},
  eprint        = {1504.00281},
  archivePrefix = {arXiv}
}

@article{Baselmans2022,
  author  = {Baselmans, J.J.A. and others},
  title   = {Ultra-sensitive {THz} microwave kinetic inductance detectors for future space telescopes},
  journal = {Astron. Astrophys.},
  year    = {2022},
  volume  = {665},
  pages   = {A17},
  doi     = {10.1051/0004-6361/202243840}
}

@ARTICLE{Dabironezare2026,
  author={Dabironezare, Shahab Oddin and Conenna, Giulia and Roos, Daan and Lamers, Dimitry and Capelo, Daniela Perez and Veen, Hendrik M. and Thoen, David J. and Anvekar, Vishal and Yates, Stephen J. C. and Jellema, Willem and Huiting, Robert and Ferrari, Lorenza and Tucker, Carole and Van Berkel, Sven L. and Day, Peter K. and Leduc, Henry George and Bradford, Charles M. and Llombart, Nuria and Baselmans, Jochem J. A.},
  journal={IEEE Transactions on Terahertz Science and Technology}, 
  title={Lens Based Kinetic Inductance Detectors With Distributed Dual Polarized Absorbers for Far Infrared Space-Based Astronomy}, 
  year={2026},
  volume={16},
  number={1},
  pages={10-26},
  doi={10.1109/TTHZ.2025.3610552}
}

@inproceedings{2022SPIE12190E..2EC,
  author    = {Cournoyer, A. and others},
  title     = {Cryogenic testing towards {TRL-5} demonstration of a novel 
               stiffness-compensated, reactionless scan mechanism for the {FTS} 
               of {SPICA} {SAFARI} instrument},
  booktitle = {Proc. SPIE},
  year      = {2022},
  volume    = {12190},
  pages     = {121902E},
  doi       = {10.1117/12.2627861}
}

@inproceedings{Cournoyer2023,
  author    = {Cournoyer, Alain and Bourque, Hugo and Carbonneau, \'Eric and Gilbert, Patrick and Houle, Simon and Boulet, Jean-Alexis and Silversides, Ian and Grandmont, Fr\'ed\'eric and Naylor, David and Gom, Brad and Christiansen, Adam and Buchan, Matthew},
  title     = {{Cryogenic Testing and Performance Characterization of a Novel Stiffness-compensated, Reactionless Scan Mechanism for a Far-infrared Post-dispersed Polarizing Fourier Transform Spectrometer}},
  booktitle = {Optica Sensing Congress 2023 (AIS, FTS, HISE, Sensors, ES)},
  pages     = {FM4B.5},
  year      = {2023},
  doi       = {10.1364/FTS.2023.FM4B.5}
}

@article{Bollinger1999,
  author  = {Bollinger, W.},
  title   = {{High precision cryogenic optics realized with ISOPHOT filter wheels}},
  journal = {Cryogenics},
  year    = {1999},
  volume  = {39},
  number  = {2},
  pages   = {149--152},
  doi     = {10.1016/S0011-2275(99)00010-7}
}

@article{pixie2011,
  author  = {Kogut, A. and others},
  title   = {The Primordial Inflation Explorer ({PIXIE}): a nulling polarimeter 
             for cosmic microwave background observations},
  journal = {J. Cosmol. Astropart. Phys.},
  year    = {2011},
  volume  = {2011},
  number  = {7},
  pages   = {025},
  doi     = {10.1088/1475-7516/2011/07/025}
}

@article{Kogut2020PIXIE,
  author  = {Kogut, A. and Fixsen, D.J.},
  title   = {Calibration method and uncertainty for the 
             Primordial Inflation Explorer ({PIXIE})},
  journal = {J. Cosmol. Astropart. Phys.},
  year    = {2020},
  volume  = {2020},
  number  = {5},
  pages   = {041},
  doi     = {10.1088/1475-7516/2020/05/041}
}

@article{Fixsen2011,
  author  = {Fixsen, D.J. and others},
  title   = {{ARCADE}~2 Measurement of the Absolute Sky Brightness at 3--90~{GHz}},
  journal = {Astrophys. J.},
  year    = {2011},
  volume  = {734},
  number  = {1},
  pages   = {5},
  doi     = {10.1088/0004-637X/734/1/5}
}

@article{Simonetto2021,
  author  = {Simonetto, A. and others},
  title   = {Millimeter-wave reflectivity tests on {MetOp-SG} {MWI} On Board Calibration Target},
  journal = {J. Instrum.},
  year    = {2021},
  volume  = {16},
  pages   = {P03036},
  doi     = {10.1088/1748-0221/16/03/P03036}
}

@inproceedings{bisou2024,
  author    = {Maffei, B. and others},
  title     = {{BISOU}: a balloon pathfinder for {CMB} spectral distortions studies},
  booktitle = {Proc. SPIE},
  year      = {2024},
  volume    = {13102},
  pages     = {131020N},
  doi       = {10.1117/12.3018371}
}

@inproceedings{TMS2020,
  author    = {Rubi{\~n}o Mart{\'\i}n, J.A. and others},
  title     = {The Tenerife Microwave Spectrometer ({TMS}) experiment: 
               studying the absolute spectrum of the sky emission in the 10--20~{GHz} range},
  booktitle = {Proc. SPIE},
  year      = {2020},
  volume    = {11453},
  pages     = {114530T},
  doi       = {10.1117/12.2561309}
}

@inproceedings{cosmo2024,
  author    = {Manzan, E. and others},
  title     = {Measuring the {CMB} spectral distortions with {COSMO}: the multimode antenna system},
  booktitle = {Proc. SPIE},
  year      = {2024},
  volume    = {13102},
  pages     = {131021A},
  doi       = {10.1117/12.3018730}
}

@article{Sauvage2025_StructuralThermalCCDR,
  author        = {Sauvage, V. and others},
  title         = {Performance characterization of a new Structural and Thermal Architecture 
                   for a future spaceborne Closed-Cycle Dilution Refrigerator},
  journal       = {arXiv preprint},
  year          = {2025},
  eprint        = {2509.21546},
  archivePrefix = {arXiv},
  primaryClass  = {astro-ph.IM}
}

@article{sauvage:2026,
  title     = {Thermal conductivity of various CFRPs from 100 mK to 20 K},
  author    = {Sauvage, Valentin},
  journal   = {arXiv},
  year      = {2026},
  doi       = {10.48550/arXiv.2602.04020}
}

@misc{Toray_T700S,
  author       = {{Toray Composite Materials America}},
  title        = {{T700S Standard Modulus Carbon Fiber: Technical Data Sheet}},
  year         = {2018},
  howpublished = {\url{https://www.toraycma.com}},
}

@inproceedings{ADR_SPIE_2026,
author = {Sylvain Martin and Anthony Attard and Florian Bancel and Ivan Charles and Jean-Louis Durand and Ad{\`e}le L{\'e}on and Christophe Martin and Adeline Robert and Elise Bellouard and Jean-Marc Duval},
title = {{5-stages ADR cooler for the Athena space mission: toward the demonstration model for X-IFU}},
volume = {14146},
booktitle = {Space Telescopes and Instrumentation 2026: Ultraviolet to Gamma Ray},
editor = {Shouleh Nikzad and Kazuhiro Nakazawa and Marco Feroci},
organization = {International Society for Optics and Photonics},
publisher = {SPIE},
pages = {141460C},
year = {2026},
doi = {10.1117/12.3104653},
URL = {https://doi.org/10.1117/12.3104653}
}

@article{fossil_general_paper,
  author      = {Aghanim, B. Maffei, B and {FOSSIL Collaboration}},
  title       = {{FOSSIL} -- {FTS} f{O}r {CMB} Spectral di{S}tortion exp{L}oration.},
  year        = {2026},
}

@article{coulon_paper_sd,
  author      = {Coulon, X and et al.},
    year = {in prep. },
}

@article{cyr_paper_sd,
  author      = {Cyr, B. and et al.},
    year = {in prep. },
}

@article{loquetlegall_paper_optics,
  author      = {Loquet Le Gall, M. and et al.},
    year = {in prep. },
}

@inproceedings{Maisonnave2022_ADR,
  author    = {Maisonnave, O. and others},
  title     = {{ADR} control electronics for {Athena}-{X-IFU} instrument, 
               status, and perspectives},
  booktitle = {Proc. SPIE},
  year      = {2022},
  volume    = {12191},
  pages     = {121912F},
  doi       = {10.1117/12.2630060}
}

@article{DUBAND2014213,
title = {SAFARI engineering model 50mK cooler},
journal = {Cryogenics},
volume = {64},
pages = {213-219},
year = {2014},
issn = {0011-2275},
doi = {https://doi.org/10.1016/j.cryogenics.2014.02.008},
url = {https://www.sciencedirect.com/science/article/pii/S0011227514000381},
author = {L. Duband and J.M. Duval and N. Luchier}
}

\end{document}